\documentclass[11pt]{article}
\usepackage{graphicx}
\usepackage{amsmath}
\usepackage{amssymb}
\usepackage{amsxtra}

\title{Creating the Universe Without a Singularity and the Cosmological Constant Problem}
\author{E. I. Guendelman\thanks{ e-mail: guendel@bgu.ac.il}\\
Physics Department, Ben Gurion University of the Negev 
\\Beer Sheva 84105, Israel} 
\begin{document}

\maketitle

\begin{abstract}
We consider a non singular origin for the Universe starting from an Einstein static Universe in the framework of a theory which uses two volume elements  $\sqrt{-{g}}d^{4}x$ and $\Phi d^{4}x$, where $\Phi $ is a metric independent density, also curvature, curvature square terms, first order formalism and for scale invariance a dilaton field $\phi$  are considered in the action. In the Einstein frame we also add a cosmological term that parametrizes the zero point fluctuations. The resulting effective potential for the dilaton contains two flat regions, for 
$\phi \rightarrow \infty$ relevant for the non singular origin of the Universe and $\phi \rightarrow -\infty$, describing our present Universe.
Surprisingly, avoidance of singularities and stability as $\phi \rightarrow \infty$ imply a positive but small vacuum energy as $\phi \rightarrow -\infty$. Zero vacuum energy density for the present universe is the "threshold" for universe creation. This requires a modified emergent universe scenario, where the universe although very old, it does have a beginning.

\end{abstract}
%   \pacs{98.80.Cq, 04.20.Cv, 95.36.+x}% PACS, the Physics and Astronomy
%                             % Classification Scheme.

%%%%%%%%%%%%%%%%%%%%%%%%%%%%%%%%%%%%%%%%%%%%%%%%%%%%%%%%%%%%%%%%%%%%%%%

The "Cosmological Constant Problem" \cite{eg2CCP1}, \cite{eg2CCP2},\cite{eg2CCP3}
(CCP), is a consequence of the  uncontrolled UV behavior of the zero point fluctuations in Quantum Field Theory (QFT), which leads to an 
equally uncontrolled vacuum energy density or cosmological constant term (CCT). This CCT is undetermined in QFT, but it is naturally very large, unless a delicate balance of huge quantities, for some unknown reason, conspires to give a very small final result. 
Also an apparently unrelated question is that of the initial condition for the inflationary universe is very important, it has been addressed for example by assuming a quantum boson condensate in the early universe\cite{eg2DK}. 

Here, we will explore a connection between the question of initial conditions for the Universe with the CCP, we will explore a candidate mechanism where the CCT is controlled, in a the context of a very specific framework, by the requirement of a non singular origin for the universe. 

We will adopt the very attractive "Emergent Universe" scenario, where conclusions concerning singularity theorems can be avoided 
\cite{eg2emerging1}, \cite{eg2emerging2}, \cite{eg2emerging3}, \cite{eg2emerging4},
\cite{eg2emerging5}, \cite{eg2emerging6}, \cite{eg2emerging7}, \cite{eg2emerging8} by  violating the geometrical assumptions of these theorems. In this scenario \cite{eg2emerging1},\cite{eg2emerging2} we start at very early times ($t \rightarrow -\infty $) with a closed static Universe (Einstein Universe).

In \cite{eg2emerging1} even models based on standard General Relativity, ordinary matter and minimally coupled scalar fields
were considered and can provide indeed a non singular (geodesically complete) inflationary universe, with a past eternal Einstein
static Universe that eventually evolves into an inflationary Universe.

Those most simple models suffer however from instabilities, associated with the instability of the Einstein static universe.
The instability is possible to cure by going away from  GR, considering non perturbative corrections to the Einstein`s field
equations in the context of the loop quantum gravity\cite{eg2emerging3}, a brane world cosmology \cite{eg2emerging4}, considering the
Starobinski model for radiative corrections (which cannot be derived from an effective action)\cite{eg2emerging5} or exotic matter\cite{eg2emerging6}. In addition to this, the consideration of a Jordan Brans Dicke model also can provide
a stable initial state for the emerging universe scenario \cite{eg2emerging7}, \cite{eg2emerging8}.

In this essay we discuss a different theoretical framework, presented in details in ref.\cite{eg2emerCCP} , 
where such emerging universe scenario is realized in a natural way, where instabilities are avoided and a successful inflationary phase
with a graceful exit can be achieved . The  model we will use was studied first in \cite{eg2SIchile} (in ref.\cite{eg2emerCCP} a few typos in \cite{eg2SIchile} have been corrected and also the discussion of some notions discussed there as well has been improved), however, we differ with \cite{eg2SIchile} in our choice of the state with (here and in ref.\cite{eg2emerCCP} with a lower vacuum energy density) that best represents the present state of the universe. This is crucial, since as it should be obvious, the discussion of the CCP depends crucially on what vacuum we take. We will express the stability and existence conditions  for the non singular initial universe in terms of the energy of the vacuum of  our candidate for the present Universe. 

We work in the context of a theory built along the lines of the two measures theory (TMT) \cite{eg2TMT1}, \cite{eg2TMT2}, \cite{eg2TMT3}, \cite{eg2TMT4} which deals with actions of the form,
\begin{equation}\label{eg2e6}
S = \int L_{1} \sqrt{-g}d^{4}x + \int L_{2} \Phi  d^{4} x    
\end{equation}

where $\Phi$ is an alternative "measure of integration", a density independent of the metric, for example in terms of four scalars $\varphi_{a}$ ($a = 1,2,3,4$),it can be  obtained as follows:

\begin{equation}\label{eg22}
\Phi =  \varepsilon^{\mu\nu\alpha\beta}  \varepsilon_{abcd}
\partial_{\mu} \varphi_{a} \partial_{\nu} \varphi_{b} \partial_{\alpha}
\varphi_{c} \partial_{\beta} \varphi_{d}
\end{equation}
and more specifically work in the context of the globally scale invariant
realization of such theories  \cite{eg2TMT2}, \cite{eg2TMT3}, which require the introduction of a dilaton field $\phi$. In the variational principle $\Gamma^{\lambda}_{\mu\nu},
g_{\mu\nu}$, the measure fields scalars
$\varphi_{a}$ and the "matter" - scalar field $\phi$ are all to be treated
as independent
variables although the variational principle may result in equations that
allow us to solve some of these variables in terms of others, that is, the first order formalism is employed, where any relation between the connection coefficients and the metric is obtained from the variational principle, not postulated a priori.
We look at the generalization of these models \cite{eg2TMT3} where an "$R^{2}$ term" is present, 
\begin{equation}\label{eg2e10}
L_{1} = U(\phi) + \epsilon R(\Gamma, g)^{2} 
\end{equation}
\begin{equation}\label{eg2e11}
L_{2} = \frac{-1}{\kappa} R(\Gamma, g) + \frac{1}{2} g^{\mu\nu}
\partial_{\mu} \phi \partial_{\nu} \phi - V(\phi)
\end{equation}
\begin{equation}\label{eg2e12}
R(\Gamma,g) =  g^{\mu\nu}  R_{\mu\nu} (\Gamma) , R_{\mu\nu}
(\Gamma) = R^{\lambda}_{\mu\nu\lambda}
\end{equation}
\begin{equation}\label{eg2e13}
R^{\lambda}_{\mu\nu\sigma} (\Gamma) = \Gamma^{\lambda}_
{\mu\nu,\sigma} - \Gamma^{\lambda}_{\mu\sigma,\nu} +
\Gamma^{\lambda}_{\alpha\sigma}  \Gamma^{\alpha}_{\mu\nu} -
\Gamma^{\lambda}_{\alpha\nu} \Gamma^{\alpha}_{\mu\sigma}.
\end{equation}
For the case the potential terms
$U=V=0$ we have local conformal invariance

\begin{equation}\label{eg2e14}
g_{\mu\nu}  \rightarrow   \Omega(x)  g_{\mu\nu}
\end{equation}

and $\varphi_{a}$ is transformed according to
\begin{equation}\label{eg2e15}
\varphi_{a}   \rightarrow   \varphi^{\prime}_{a} = \varphi^{\prime}_{a}(\varphi_{b})
\end{equation}

\begin{equation}\label{eg2e16}
\Phi \rightarrow \Phi^{\prime} = J(x) \Phi    
\end{equation}
 where $J(x)$  is the Jacobian of the transformation of the $\varphi_{a}$ fields.

This will be a symmetry in the case $U=V=0$ if 
\begin{equation}\label{eg2e17}
\Omega = J
\end{equation}
Notice that $J$ can be a local function of space time, this can be arranged by performing for the 
$\varphi_{a}$ fields one of the (infinite) possible diffeomorphims in the internal $\varphi_{a}$ space.

In the case we have potentials non zero $U$ and $V$, we give up local conformal invariance, but can retain global scale invariance which is satisfied if \cite{eg2TMT3}, \cite{eg2TMT2}($ f_{1}, f_{2},\alpha $ being constants),
\begin{equation}\label{eg2e19} 
V(\phi) = f_{1}  e^{\alpha\phi},  U(\phi) =  f_{2}
e^{2\alpha\phi}
\end{equation}

Notice that in this way we have chosen all the conformal breaking to be through the potential, the kinetic terms do not break conformal invariance. In this sense the breaking of conformal invariance is what is usually called a soft breaking. Consideration of cosmological models   (in particular emergent models) with "non soft" breaking of conformal invariance has been considered also \cite{eg2CGKHL}.
A particularly interesting equation is the one that arises from the $\varphi_{a}$
fields, this yields $L_2 = M$, where $M$ is a constant that spontaneously breaks scale invariance.
The 
Einstein frame,  which is a redefinition of the metric by a conformal factor, is defined as 

\begin{equation}\label{eg2e47}
\overline{g}_{\mu\nu} = (\chi -2\kappa \epsilon R) g_{\mu\nu}
\end{equation}

where $\chi$ is the ratio between the two measures, $\chi =\frac{\Phi}{\sqrt{-g}}$, determined from the consistency of the equations to be $\chi = \frac{2U(\phi)}{M+V(\phi)}$. The relevant fact is that the connection coefficient equals the Christoffel symbol of this new metric (for the original metric this "Riemannian" relation does not hold). There is a "k-essence" type effective action, where one can use this Einstein frame metric. As it is standard in treatments 
 of theories with non linear kinetic terms or k-essence models\cite{eg2k-essence1}-\cite{eg2k-essence4}, it is determined by a pressure functional, 
($X = \frac{1}{2} \overline{g}^{\mu\nu}\partial_{\mu} \phi \partial_{\nu} \phi $).

\begin{equation}
S_{eff}=\int\sqrt{-\overline{g}}d^{4}x\left[-\frac{1}{\kappa}\overline{R}(\overline{g})
+p\left(\phi,R\right)\right] \label{eg2k-eff}
\end{equation}

\begin{equation}
 p = \frac{\chi}{\chi -2 \kappa \epsilon R}X - V_{eff}
\end{equation}
where $V_{eff}$ is an effective potential for the dilaton field given by

\begin{equation}\label{eg2e50}
 V_{eff}  = \frac{\epsilon R^{2} + U}{(\chi -2 \kappa \epsilon R)^{2} }
\end{equation}

$\overline{R}$ is the Riemannian curvature scalar built out of the bar metric, $R$ on the other hand is the non Riemannian curvature scalar defined in terms of the connection and the original metric,which turns out to be given by $R = \frac{-\kappa (V+M) +\frac{\kappa}{2} \overline{g}^{\mu\nu}\partial_{\mu} \phi \partial_{\nu} \phi \chi}
{1 + \kappa ^{2} \epsilon \overline{g}^{\mu\nu}\partial_{\mu} \phi \partial_{\nu} \phi}$. This $R$ can be inserted in the action (\ref{eg2k-eff})
or alternatively, $R$ in the action (\ref{eg2k-eff}) can be treated as an independent degree of freedom, then its variation gives the required value
as one can check (which can then be reinserted in (\ref{eg2k-eff})).
Introducing this $R$ into the expression (\ref{eg2e50}) and considering a constant field $\phi$ we find that $ V_{eff}$ has two flat
regions. The existence of two flat regions for the potential
is shown to be consequence of the s.s.b. of the scale symmetry (that is of considering  $M \neq 0$ ).  
The quantization of the model can proceed from (\ref{eg2k-eff}) (see discussion in \cite{eg2emerCCP}) and additional terms could
be generated by radiative corrections. We will focus only on a possible cosmological term in
the Einstein frame added (due to zero point fluctuations) to (\ref{eg2k-eff}), which leads then to the new action
\begin{equation}
S_{eff,\Lambda }=\int\sqrt{-\overline{g}}d^{4}x\left[-\frac{1}{\kappa}\overline{R}(\overline{g})
+p\left(\phi,R\right)- \Lambda \right] \label{eg2act.lambda}
\end{equation}

This addition to the effective action leaves the equations of motion of the scalar field unaffected, but the gravitational equations acquire a 
cosmological constant. Adding the $\Lambda$ term can be regarded as a redefinition of $V_{eff}\left(\phi,R\right)$
\begin{equation}
V_{eff}\left(\phi,R\right) \rightarrow V_{eff}\left(\phi,R\right) + \Lambda  \label{eg2V.lambda}
\end{equation}

In this resulting 
model, there are two possible types of emerging universe solutions,
for one of those, the initial Einstein Universe (realized in the region $\phi \rightarrow \infty$ ) can be stabilized due to the nonlinearities of the model, if $\epsilon<0$, $f_2>0$  and $f_2 + \kappa^2 \epsilon f^2_1>0$ provided the vacuum energy 
density of the ground state, realized in the region $\phi \rightarrow -\infty$, being given by 
$V_{eff}\rightarrow \frac{1}{4\epsilon \kappa ^{2}} + \Lambda = \Delta \lambda$  is  positive, but not very large, since it should be bounded from above by 
the inequality $\Delta \lambda < \frac{1}{12(-\epsilon)\kappa^2 } \left[ \frac{f_2}{f_2 + \kappa^2 \epsilon f^2_1 }  \right]$. These are very satisfactory results, since it means that the existence and stability
of the emerging universe prevents the vacuum energy in the present universe from being very large, but requires it to be positive. The transition from the emergent universe to the ground state goes through an intermediate inflationary phase, therefore reproducing the basic standard cosmological model as well.
So, it turns  out that the creation of the universe can be considered as a "threshold event" for zero present vacuum energy density, which naturally gives a positive but small vacuum energy density for the present universe.

One may ask the question: how is it possible to discuss the "creation of the universe" in the context of the "emergent universe"?. After all, the Emergent Universe basic philosophy is that the  universe had a past of infinite duration. However, that most simple notion of an emergent universe with a past of infinite duration has been recently challenged by
Mithani and Vilenkin \cite{eg2vilenkin}, at least in the context of a special  model. They have shown that a completely stable emergent universe, although completely stable classically, could be unstable under a tunneling process to collapse. On the other hand, an emergent universe can indeed be created by a tunneling process as well.

After creation from "nothing" an emerging universe could last for a long time, provided it is classically stable, that is where the constraints on the cosmological constant for the late universe discussed here come in. If it is not stable, the emergent universe will not provide us with an appropriate "intermediate state" connecting the creation of the universe with the present universe. The existence of this stable intermediate state provides in our picture the reason for the universe to prefer a very small vacuum energy density at late times, since universes that are created, but do not make use of the intermediate classically stable emergent universe will almost immediately recollapse, so they will not be "selected".

The situation is somewhat similar to the reason Carbon is formed at a reasonable rate in stars\cite{eg2Hoyle}. There, it is also an appropriate resonance that makes the creation of carbon possible. The analogous role of that resonance, when refering to the creation of the whole universe are the stable emergent universe solution in the picture we are considering and 
instead of carbon, the "product" we are trying to explain is a small cosmological constant.

\section*{Acknowledgements}

It is a pleasure for me to thank Norma Manko\v c, Holger Nielsen, Maxim Yu. Khlopov  and all the organizers and participants of the Bled workshop "On What Comes Beyond The Standard Models?" for a very productive and  interesting conference. I would also like to thank Alexander Vilenkin for correspondence concerning his work on instability and possibility of quantum creation of an emergent universe.

\end{document}